\documentstyle[12pt]{article}
\newcommand{\la}[1]{\label{#1}}
\newcommand{\be}{\begin{equation}}
\newcommand{\ee}{\end{equation}}
\newcommand{\ba}{\begin{eqnarray}}
\newcommand{\ea}{\end{eqnarray}}
\newcommand{\bi}{\begin{itemize}}
\newcommand{\ei}{\end{itemize}}
\newcommand{\rmi}[1]{{\mbox{\scriptsize #1}}}

\newcommand{\nr}[1]{(\ref{#1})}
\newcommand{\tr}{{\rm Tr\,}}
\newcommand{\nn}{\nonumber \\}

\renewcommand{\vec}[1]{{\bf #1}}
\newcommand{\eq}{Eq.~}
\newcommand{\eqs}{Eqs.~}

\def\lsi{\raise0.3ex\hbox{$<$\kern-0.75em\raise-1.1ex\hbox{$\sim$}}}
\def\gsi{\raise0.3ex\hbox{$>$\kern-0.75em\raise-1.1ex\hbox{$\sim$}}}

\begin{document}

\begin{titlepage}
\begin{flushright}
CERN-TH/99-206\\
UNIL-IPT/99-2\\
hep-th/9907194\\
\end{flushright}
\begin{centering}
\vfill

{\bf AN ALL-ORDER DISCONTINUITY AT\\ THE ELECTROWEAK PHASE TRANSITION}
\vspace{0.8cm}

M. Laine$^{\rm a,b,}$\footnote{mikko.laine@cern.ch}, 
M. Shaposhnikov$^{\rm c,}$\footnote{mikhail.shaposhnikov@ipt.unil.ch} \\

\vspace{0.3cm}
{\em $^{\rm a}$Theory Division, CERN, CH-1211 Geneva 23,
Switzerland\\}
\vspace{0.3cm}
{\em $^{\rm b}$Dept.\ of Physics,
P.O.Box 9, FIN-00014 Univ.\ of Helsinki, Finland\\}
\vspace{0.3cm}
{\em $^{\rm c}$Institute for Theoretical Physics, University of  
Lausanne,\\ 
BSP-Dorigny, CH-1015 Lausanne, Switzerland}

\vspace{0.7cm}
{\bf Abstract}

\end{centering}

\vspace{0.3cm}\noindent 

We define a non-local gauge-invariant Green's function which can
distinguish between the symmetric (confinement) and broken (Higgs) phases 
of the hot SU(2)$\times$U(1) electroweak theory to all orders in the 
perturbative expansion. It is related to the coupling of the Chern-Simons 
number to a massless Abelian gauge field. The result implies either that 
there is a way to distinguish between the phases, even though the macroscopic 
thermodynamical properties of the system have been observed to 
be smoothly connected, or that the perturbative Coleman-Hill theorem on 
which the argument is based, is circumvented by non-perturbative effects. 
We point out that this question could in principle be studied 
with three-dimensional lattice simulations. 

\vfill

\noindent
CERN-TH/99-206\\
UNIL-IPT/99-2\\
August 1999

\vfill

\end{titlepage}

\paragraph{Introduction.}
Non-Abelian gauge theories based on groups SU($N$),
$N \geq 2$, with scalar fields in the fundamental representation have no
known gauge-invariant order parameters that can distinguish between the 
symmetric (confinement) and broken (Higgs) phases~\cite{elitzur}. 
This leads to the conclusion that the confinement phase 
and the Higgs phase can be smoothly
connected~\cite{fradkin}. In other words, there may be no separate Higgs
and confinement phases but just one Higgs-confinement phase. This
consideration tells that finite temperature phase transitions in
spontaneously broken
gauge theories are not a generic phenomenon: depending on the
parameters of the theory the system may go through a first order
phase transition (or, in special cases, a second order one), 
but it may also be that there is no phase transition at all.

These expectations have been confirmed in lattice Monte Carlo
simulations. In particular, no phase transition is observed
for sufficiently large scalar self-coupling for
three-dimensional (3d) theories like SU(2) with a doublet of 
Higgses~\cite{Kajantie:1996mn}. As 3d theories can be considered as
the effective theories of high temperature four-dimensional (4d) theories 
these results mean that there is no finite temperature phase transition 
in these systems in some region of the parameter space (for Higgs boson
masses larger than the W boson mass).

The behaviour of an Abelian gauge theory is different. Although there is
no local gauge-invariant order parameter that can distinguish between
the Coulomb and the Higgs phases, there are non-local ones, such as
the photon mass. In the Higgs phase the photon mass is non-zero, while in
the Coulomb phase it is zero, clearly indicating a difference between the
phases. This statement is correct in all orders of perturbation
theory, and it has also been tested in lattice simulations~\cite{masses}.
There are also other non-perturbative order parameters in the Abelian case, 
such as the tension of an infinitely long vortex~\cite{Kajantie:1998zn}.

Here we are interested in theories that contain both Abelian and
non-Abelian parts and contain scalar fields in the fundamental
representation. We will deal essentially with the electroweak theory
(or some typical extensions thereof such as the MSSM, 
with a similar pattern of electroweak symmetry breaking),
although the results remain valid for many models of the type $G\times$U(1)
where $G$ is a simple group. As in both previous cases, there are no
local gauge-invariant order parameters that can select the symmetric or
the Higgs phase. A non-local order parameter
associated with the existence of a massless vector excitation does not
distinguish between the phases either, as the photon in the
Higgs phase and the hypercharge vector boson in the symmetric phase
remain massless in all orders of perturbation theory (this statement
has also been checked with
non-perturbative 3d lattice simulations~\cite{Kajantie:1997qd}). 
The projection of the massless 
state to the gauge-invariant Abelian hypercharge
field strength was found to be a
smooth function of the temperature at sufficiently large Higgs masses
\cite{Kajantie:1997qd}, and all the other observables measured
(expectation values of local gauge-invariant operators; other
masses) also behaved smoothly. 
Thus, it seems that the confinement and Higgs
phases can be smoothly connected also for the SU(2)$\times$U(1) theory
and that there may be no phase transition.

In this paper we define a non-local 
Green's function which nevertheless jumps when one
goes from the symmetric to the Higgs phase {\em independently} of 
the scalar self-coupling (or the Higgs mass) to all orders in 
perturbation theory.  The existence of a massless vector
excitation (or, in other words, the presence of an unbroken Abelian
gauge symmetry) in both phases of the electroweak theory is essential for the
argument. Therefore, one {\em can} 
separate the confinement phase from the Higgs
phase in the SU(2)$\times$U(1) gauge theory to all orders
in the perturbative expansion. It remains to be seen if this
statement is valid beyond perturbation theory.

\paragraph{Phase transition in 3d with a topological mass term.} 
Let us consider the 3d
SU(2)$\times$U(1) theory with a single Higgs doublet. This theory is
the high temperature limit for the standard 4d electroweak
theory and many of its extensions, 
obtained by integrating out fermions, the non-zero 
Matsubara modes of bosons, and the zero Matsubara modes
of the temporal components of the gauge fields~\cite{generic}. 
Below we will see how the same results can be derived directly in 4d.
The 3d action is
\be
S  =  \int\! d^3x \biggl\{\frac{1}{4}B_{ij}B_{ij}+
\frac{1}{2}\tr F_{ij}F_{ij}+
(D_i\phi)^{\dagger}(D_i\phi)+
m_3^2\phi^{\dagger}\phi+\lambda_3
(\phi^{\dagger}\phi)^2 \biggr\},
\label{lagr}
\ee
where $F^a_{ij}=\partial_iA_j^a-\partial_jA_i^a+
g\epsilon^{abc}A^b_iA^c_j$, $F_{ij}=T^a F^a_{ij}$,
$D_i\phi=(\partial_i-ig A_i+i(g'/2)B_i)\phi$, $A_i=T^a A_i^a$,
$B_{ij}=\partial_iB_j-\partial_jB_i$ and $T^a=\tau^a/2$ (the $\tau^a$
are the Pauli matrices). The 3d gauge couplings $g,g'$ have
the dimension GeV$^{1/2}$.  

Let us first consider a somewhat more complicated theory by adding to
\eq\nr{lagr} topological mass terms \cite{topmass} for the U(1)  
and SU(2) fields: $S\to S+\Delta S$, where
\ba 
\Delta S & = & i \Bigl( \mu_1 N_\rmi{CS}^{(1)} - \mu_2 N_\rmi{CS}^{(2)}\Bigr),
\label{top} \\
N_\rmi{CS}^{(1)} & = &  \frac{g'^2}{16 \pi}
\int\! d^3x \epsilon_{ijk} B_{ij} B_k, \\
N_\rmi{CS}^{(2)} & = & \frac{g^2}{8 \pi}
\int\! d^3x \epsilon_{ijk}\tr \left(F_{ij} A_k + \frac{2}{3} i g 
A_i A_j A_k\right).
\ea
The theory with a topological mass term for 
the SU(2) field is mathematically
consistent beyond perturbation theory only if $\mu_2$ is quantized
\cite{topmass}, $\mu_2 = 0,\pm 1, \pm 2, ...$, while the topological
mass for the U(1) field may be arbitrary. In the following we
will nevertheless take $\mu_1 = \mu_2 =\mu$:
it is in this case that we find a discontinuity computable to 
all orders in perturbation theory, and this is also the limit
which has an interpretation in the 4d finite temperature 
electroweak theory (see below).

Consider now the vacuum polarization tensor 
(the inverse propagator) for the U(1) field. 
In momentum representation, in a general 
covariant gauge with the gauge parameter $\xi$, 
it is given by
\be
G_{ij}^{-1}(k,\mu) = (k^2 \delta_{ij} - {k_i k_j})\Pi_1(k^2,\mu) +
i \epsilon_{ijl} k_l \Pi_2(k^2,\mu) + 
\xi^{-1} k_i k_j, 
\label{green} 
\ee 
where 
\be
G_{ij} (k,\mu) = \int d^3\! x\, e^{ikx}\langle B_i(x)B_j(0)\rangle.
\la{propag}
\ee
The parity odd part $\Pi_2(k^2,\mu)$ is non-vanishing for $\mu\neq 0$, 
and gauge-independent. We will 
be interested in the limit $\Pi_2(k^2\to 0,\mu)$.

Consider first $\Pi_2(0,\mu)$ in the symmetric phase.
At the lowest order of perturbation theory, 
it is proportional to the topological mass,
\be
\Pi_2(0,\mu) = - i \mu \frac{g'^2}{4\pi}.
\label{tree}
\ee 
However, the theory under consideration satisfies all the 
requirements of the Coleman-Hill (CH) 
theorem~\cite{Coleman:1985zi}: there is an unbroken
Abelian gauge symmetry and no massless charged excitations
(that is, 
as long as we stay away from the transition point $m_3^2\approx 0$;
the result however does not depend on $m_3^2$). 
Thus, all corrections to $\Pi_2(0,\mu)$
vanish beyond 1-loop level. 
Consequently, $\Pi_2(0,\mu)$ can be computed exactly at the 1-loop level. 
As is easy to
understand, there are in fact no 1-loop corrections either,
because on the 1-loop level the vacuum polarization tensor is given by
scalar loops which contain no trace of the topological mass term. 
Thus, in the symmetric phase the relation in \eq\nr{tree} is
valid to all orders in perturbation theory.

Let us then compute $\Pi_2(k^2,\mu)$ (deep) in the Higgs phase. 
We are interested in the limit
$k^2 \rightarrow 0$, so that
only external lines associated
with what would be massless particles in the absence of 
the topological mass term are important. We thus concentrate
on the electromagnetic field $Q_i=-\sin\!\theta A^3_i + \cos\!\theta B_i$ 
where $\tan\!\theta = g'/g$,  
and the idea is to apply again the CH theorem, but now for $Q_i$.
One of the conditions of the theorem is
actually not valid: there is a bilinear mixing between the massive 
field $Z_i= \cos\!\theta A^3_i + \sin\!\theta B_i$
and the photon field $Q_i$ of the form 
$\epsilon_{ijk} Z_i \partial_j Q_k$. 
However, this mixing can be
formally removed 
(i.e., the quadratic part of the Lagrangian can be diagonalized)
by a shift of $Z_i$:
in momentum space,   
\ba 
Z_i(k) & \rightarrow  &  Z_i(k) + 
\biggl[ 
\Bigl(\delta_{ij} - \frac{k_ik_j}{k^2}\Bigr) A(k^2) + 
2 i \epsilon_{ijl} k_l B(k^2)  
\biggr] Q_j(k), \nn
A(k^2) & = & \frac{8 k^2}{k^2+m_Z^2} 
\biggl( \frac{g^2+g'^2}{16\pi} \mu \biggr)^2
\Delta^{-1}(k^2) \cos\! 2\theta \sin\! 2\theta, \nn
B(k^2) & = & 
i \biggl(
\frac{g^2+g'^2}{16\pi} \mu \biggr)
\Delta^{-1}(k^2) \sin\! 2\theta, \nn 
\Delta(k^2) & = &  
k^2 + m_Z^2 +\frac{k^2}{k^2+m_Z^2} 
\biggl(
\frac{g^2+g'^2}{4 \pi} \mu \cos\! 2\theta
\biggr)^2.
\label{shift}
\ea 
This brings the effective 
Lagrangian in a form consistent with the CH conditions.
This transformation is non-local, 
but it is infrared insensitive and does not break the 
U(1) gauge invariance of $Q_i$, 
and thus does not invalidate the CH-theorem. 

As a result of writing the action in terms of $Q_i,Z_i$
in the Higgs phase
and making the transformation
in \eq\nr{shift}, there is a  topological mass
term for the field $Z_i$, but none for $Q_i$ \footnote{In fact, 
one does generate a parity-odd term $\sim \epsilon_{ijk}Q_i\partial_j Q_k$ 
proportional to $\mu^3 k^2/m_Z^4$, 
but this does not contribute in the limit $k^2\to 0$ we are 
interested in.}. Thus, 
in the vacuum polarization tensor for
the massless field $Q_i$, 
\be 
\Pi_2^{(Q)}(0,\mu) = 0
\ee  
at the tree-level. 
An explicit computation of 1-loop diagrams to order $\mu$ 
gives zero for $k^2\to0$ 
(we have
carried out this computation before making 
the shift in \eq\nr{shift}), 
and higher order corrections are absent due to the CH theorem.  
Thus, the values of $\Pi_2(0,\mu)$ associated with a massless field
are different in the symmetric and
Higgs phases, and this is valid in all orders of perturbation theory.

The fact that the IR-sensitive part of the effective action shows
such behaviour, suggests that the partition function of the system
(the minimum of the effective action) may also behave
non-analytically at the transition point for $\mu\neq 0$.  In other
words, it is likely that this theory exhibits a genuine first or
second order phase transition independent of the value of the
scalar self-coupling: there are no massless excitations in the
symmetric phase, but there is a massless particle in the Higgs phase.

\paragraph{Phase transition in 3d without topological mass term.} 
The result obtained above
can be applied to the original 3d theory in \eq\nr{lagr}, 
without the
topological mass term. Note that in perturbation theory,
the quantization of
$\mu$ is not essential and the statement about the absence of higher
order corrections to $\Pi_2(0,\mu)$ is valid for real values of~$\mu$. 
Thus, one
can consider the derivative with respect to $\mu$ of 
$G^{-1}_{ij}(k,\mu)$
for the theory with $\mu \neq 0$, and then take $\mu=0$. The
Green's function that results from this operation is given by
\be
D_{ij}(k)= 
G_{il}^{-1}(k,0) S_{lm}(k)G_{mj}^{-1}(k,0),
\label{def}
\ee
where
\be 
S_{ij}(k)=\int d^3x e^{ikx}\Bigl \langle B_i(x) B_j(0)  
\Bigl(N_\rmi{CS}^{(1)}-N_\rmi{CS}^{(2)}\Bigr)\Bigr\rangle
\label{order} 
\ee 
is a Green's function computed in the theory with 
$\mu=0$ \footnote{To avoid confusion 
let us note that this Green's function is
not related to the rate of sphaleron transitions which was studied
numerically in the crossover region in the context of the 
pure SU(2)+Higgs model in~\cite{sd}.}.
The statement is that in the symmetric phase,
\be
D \equiv 
\lim_{k^2\to 0} D_{ij}(k) \frac{\epsilon_{ijl} k_l}{k^2} = 
\left. 2 \frac{d}{d\mu} \Pi_2(0,\mu) \right|_{\mu=0} = 
- i \frac{g'^2}{2\pi} \hspace*{0.5cm} (\mbox{\rm symmetric phase}). \la{op}
\ee
This is valid in all orders of perturbation theory. 

In the Higgs phase, in contrast, $D=0$.
As we have argued above,  
the massless field $Q_i$
does not couple to $N_\rmi{CS}^{(1)}-N_\rmi{CS}^{(2)}$
in any order of perturbation theory for $k^2\to 0$, 
so that $S_{ij}(k)$ in \eq\nr{order} does not have a term behaving as
$\epsilon_{ijl}k_l/k^4$.
At the same time, the $G_{ij}^{-1}(k,0)$ appearing in 
\eq\nr{def} and defined in \eq\nr{propag},
still behaves as $\propto k^2$ for $k^2\to 0$.
This leads to the vanishing of $D$ in the Higgs phase.

A subtle point in this argument is that the quantity obtained from
\eq\nr{op} is gauge-invariant only with respect to topologically
trivial (small) SU(2) gauge transformations. To make the order parameter
completely gauge-invariant one should replace $N_\rmi{CS}^{(2)}$ by 
some operator which is completely gauge-invariant but coincides with 
$N_\rmi{CS}^{(2)}$ 
in perturbation theory.  Such an operator can be defined 
in the 4d theory with chiral fermions, see below. 
Here we recall a definition
within the 3d theory~\cite{mes,Ambjorn:1989gf}
which can be successfully discretized and implemented in practical 
lattice simulations~\cite{measurement}.

The idea is to replace $N_\rmi{CS}^{(2)}$ with the 
difference 
$N_\rmi{CS}^{(2)}-N_\rmi{CS,cl}^{(2)}$, 
where $N_\rmi{CS,cl}^{(2)}$ is a particular integer times $2\pi$
depending on the initial field configuration, with 
the properties that  
$N_\rmi{CS}^{(2)}-N_\rmi{CS,cl}^{(2)}$ is gauge-invariant 
and that
$N_\rmi{CS,cl}^{(2)}$ vanishes in case the initial field
configuration is ``perturbative'' (i.e., $g A^2 \ll \partial A$). 
Such a difference can be obtained with the following algorithm.
Suppose we have some (sufficiently smooth)
3d gauge field configuration. Introduce a fictitious
time variable $\tau$ and ``cool'' the
configuration according to 
\be
\frac{\partial \Phi}{\partial \tau} = - \frac{\delta S}{\delta \Phi},
\ee
where $\Phi$ is a generic notation for all real field
components. The cooling, which is a gauge covariant procedure
for static gauge transformations, 
is to be continued all the way from $\tau=0$
to a classical vacuum configuration at $\tau\to\infty$.
Define the non-Abelian and Abelian electric fields as $E_i^B = \frac{\partial
B_i}{\partial \tau}, E_i^a = \frac{\partial A_i^a}{\partial \tau}$,
compute the topological number densities $\epsilon_{ijk}E_i^B B_{jk}$  
and $\epsilon_{ijk}E_i^a F_{jk}^a$, and 
integrate the difference of them over space and
the fictitious time $\tau$. This gives a gauge-invariant result
$N_\rmi{CS}^{(2)}-N_\rmi{CS,cl}^{(2)}$, 
which can replace the SU(2) Chern-Simons number in \eq\nr{order}:
if the initial configuration is perturbative, the system
should behave essentially as in the Abelian case, so that
indeed $N_\rmi{CS,cl}^{(2)}=0$.

For a completely gauge-invariant measurement one should also
replace $B_i$ by the hypermagnetic field
$\epsilon_{ijk} B_{jk}$ in \eqs\nr{propag}, \nr{order}\footnote{
In principle, one can then relate the Green's function
in \eq\nr{order} to the 
expectation value 
$\langle( N_\rmi{CS}^{(1)}-N_\rmi{CS}^{(2)})\rangle$,
measured in an appropriate inhomogeneous external hypermagnetic field.}. 
Note that for the measurement of $G_{ij}(k,0)$, 
$N^{(2)}_\rmi{CS}$ is not needed.

\paragraph{Phase transition in 4d.} 
The consideration above
can be carried out also directly in the finite temperature 4d theory. Let
us take  as an example the standard electroweak theory, 
with its real fermionic
content.  Consider the finite temperature  
(one can also have a finite chemical potential for {\em conserved}
charges)
Green's
function defined in \eq\nr{def}, but now in the 4d theory and with  
the replacement of $S_{ij}$ by 
\be 
S_{ij}(k)=\int d^3x e^{ikx} \langle B_i(x) B_j(0) ({B+L})\rangle,
\label{ferm}   
\ee
where ${B+L}=\int d^3x j_0^{B+L}$ and
$j_{\mu}^{B+L}$ is the baryon + lepton number current, 
generated by the global transformation 
$\psi_i \to \exp(i\alpha)\psi_i$. 
We will be interested in the limit
$k^0=0, \vec{k} \rightarrow 0$. 
The part of $S_{ij}(k)$ antisymmetric in $i,j$
contains a term linear in $\vec{k}$, 
precisely as in \eq\nr{op}. Now, the proof of the CH theorem 
(the absence of renormalization of the term linear in $\vec{k}$ beyond 
1-loop level) in fact does not essentially 
depend on the number of dimensions
(as long as there are no new IR-problems) and is
directly applicable to our case. 
An explicit 1-loop computation gives
a non-zero value in the symmetric phase since the Abelian hypercharge
field contributes to the anomaly in the baryonic current. At the same
time, an explicit computation in the Higgs phase gives zero, as there
is no contribution to the baryonic current anomaly coming exclusively from
the electromagnetic field. This statement does not depend on the mass 
spectrum of the theory, nor on whether there is a finite temperature $T$,
or a finite chemical potential $\mu$, with the associated
$Z_0,Q_0$-condensate (this does not break the 
spatial U(1)-invariance).
Thus, the object defined in \eq\nr{ferm}
can serve as a distinction between the confinement and Higgs 
phases to all orders in perturbation theory. 

In the 4d SU(2)$\times$U(1) theory without fermions 
one could use again the Chern-Simons number as in \eq\nr{order},  
instead of the baryon + lepton number. 

\paragraph{Conclusions.} 
We proposed a gauge-invariant Green's function which
can distinguish between
the symmetric and Higgs phases of the electroweak
theory. It is related to Chern-Simons number of the gauge fields and
has a different value in the two phases, computable to all
orders in perturbation theory. The most important ingredients of our
consideration are the existence of an unbroken 
Abelian U(1) group in the symmetric
and Higgs phases, the non-trivial photon mixing in the Higgs phase, 
the absence of massless charged particles, 
and the use of the Coleman-Hill theorem. It remains to be seen whether 
the CH theorem is valid beyond perturbation theory
in the present context. 

Obviously, this consideration does not allow us to fix the parameters
of the theory at which the jump of the Green's function takes place, 
precisely as in the consideration of the photon mass in the case 
of the Abelian Higgs
model. Nevertheless, if non-perturbative effects were absent, 
it would tell
that there is a non-analytic behaviour in
a certain Green's function when
one goes from the symmetric to the Higgs phase. This
does not as such necessarily mean that the vacuum energy 
of the 3d theory (or the partition function of the finite 
temperature 4d theory) has any singularity at the point of the
transition, since the Chern-Simons number does not appear 
as a term in the Lagrangian for $\mu=0$
and serves only as an external probe.

There is a physical consequence from this result. Suppose that one has
a high temperature system with a non-zero chemical potential $\mu_\rmi{$B+L$}$ 
for the baryon + lepton number. This system is unstable as $B+L$
is not conserved due to the anomaly, 
but can nevertheless be described on the 
perturbative level by an effective action of the type in \eqs\nr{lagr}, 
\nr{top}, with the replacement $\mu \to -i \mu_\rmi{$B+L$}$~\cite{cs}. 
Now, the resulting system is unstable even perturbatively in the symmetric 
phase due to the coupling of the Chern-Simons number to the massless 
hypercharge vector field, which leads to a spontaneous generation of a
hypermagnetic field \cite{Joyce:1997uy}. 
However, in the Higgs phase no magnetic field is generated
perturbatively, due to the absence of a coupling
of $N_\rmi{CS}$ to a massless mode, as we have 
discussed above. 


\vspace*{0.5cm}

We thank K. Kajantie, 
S. Khlebnikov and K. Rummukainen for many helpful discussions.
The work of 
M.L. was partly supported by the TMR network {\em Finite Temperature Phase
Transitions in Particle Physics}, EU contract no.\ FMRX-CT97-0122.
The work of M.S. was partly supported by the Swiss Science Foundation,
contract no. 21-55560.98.

\end{document}